# Analytic Fourier ptychotomography for volumetric refractive index imaging

Zhenyu Dong[1,†], Haowen Zhou[1,†], Ruizhi Cao[1,†,*], Oumeng Zhang[1], Shi Zhao[1], Panlang Lyu[1], Reinaldo Alcalde[2], and Changhuei Yang[1]

[1]Department of Electrical Engineering, California Institute of Technology, Pasadena, CA, USA
[2]Division of Biology and Biological Engineering, California Institute of Technology, Pasadena, CA, USA
[†]These authors contributed equally to this work.
[*]rcao@alumni.caltech.edu

## Abstract

Three-dimensional (3D) refractive index (RI) tomography offers label-free, quantitative volumetric imaging but faces limitations due to optical aberrations, limited resolution, and computational complexity inherent to existing approaches. To overcome these barriers, we propose Analytic Fourier Ptychotomography (AFP), a new computational microscopy technique that analytically reconstructs aberration-free, complex-valued 3D RI distributions without iterative optimization or axial scanning. AFP incorporates a new concept named finite sample thickness (FST) prior, analytically solving the inverse scattering problem through three sequential steps: complex-field reconstruction via the Kramers-Kronig relation, linear aberration correction using overlapping spectra, and analytic spectrum extension into the darkfield region. Unlike iterative reconstruction methods, AFP does not require parameter tuning and computationally intensive optimizations – which are error-prone and non-generalizable. We experimentally demonstrated that AFP significantly enhances image quality and resolution under various aberration conditions across a range of applications. AFP corrected aberrations associated with 25 Zernike modes (with maximal phase difference of $2.3\pi$ and maximal Zernike coefficient value of 4), extended the synthetic numerical aperture from 0.41 to 0.99, and provided a two-fold resolution enhancement in all directions. AFP's simplicity and robustness makes it an attractive imaging technology for quantitative 3D analysis in biological, microbial ecological, and medical studies.

## Introduction

Three-dimensional (3D) refractive index (RI) imaging, or RI tomography[1,2], is a relatively new class of microscopy methods. RI imaging is label-free, quantitative, and capable of elucidating biological structures at high resolution. Unlike traditional through-focus scanning methods which acquire qualitative absorption or phase contrast images layer by layer, RI tomography computationally reconstructs the 3D RI distribution from multiple measurements taken at different illumination angles by a planar sensor. The use of angular illumination scan eliminates the need for precise and repeatable axial mechanical scanning and, thus, significantly improves system robustness and cuts down on the required scan time. RI tomography is an attractive alternative to fluorescence microscopy, the dominant life science microscopy modality, in that it does not require the use of fluorescence agents and is, thus, much more resilient against phototoxicity – a common fluorescence microscopy challenge especially when live samples are involved.

The field of RI tomography has seen some rapid and impactful developments in the form of innovative implementation strategies. However, each of these methods is associated with its own particular disadvantages. For example, optical diffraction tomography[3,4] utilizes high-coherence lasers and interferometric measurements to obtain complex light fields, enabling 3D RI imaging. However, an interferometric setup results in complicated system configurations, and the use of high-coherence lasers are associated with speckle noise. Non-interferometric RI imaging approaches have been developed to address this. For example, techniques such as 3D differential phase contrast microscopy[5], intensity diffraction tomography[6,7], and low-coherence holotomography[2,8,9] utilize transfer function theory alongside deconvolution methods to reconstruct RI volumes without the use of inteferometric setups. However, these methods require careful parameter tuning and regularization to obtain reliable RI reconstructions. Similarly, optimization-based methods[10–16], such as multi-slice[10,11] and multi-layer Born[12], are sensitive to initial conditions, require precise parameter selection, and are computationally demanding. To overcome these limitations, analytic methods based on the Kramers-Kronig (K-K) relation have been proposed for 3D RI imaging, leveraging either numerical aperture (NA)-matching illuminations[17] or a combination of through-focus and multi-angle brightfield acquisitions[18], to enable improved reconstruction stability. However, these analytic methods either suffer from limited axial resolution[17] or require axial scanning and significantly more measurements[18].

Besides these method-specific issues, there are two major challenges that persist in the field of RI tomography. The first is the aforementioned resolution limitation imposed by the NA of the objective lens. Fairly recently, Fourier ptychographic diffraction tomography[19–21], which is derived from Fourier ptychographic microscopy in two-dimensional (2D) imaging[22,23], has attempted to address this limitation. It reconstructs the complex light field by numerically stitching together sample's spectra from multi-angle illuminations, including darkfield images that contain higher spatial frequencies. However, it still relies on iterative optimization techniques to address the non-convex phase retrieval problem, inheriting the aforementioned limitations of optimization-based methods. The second major challenge associated with RI tomography is optical aberration, which results from imperfections in optical elements (particularly at certain wavelengths such as ultraviolet), system misalignment, and sample RI mismatches. These imperfections can distort the original wavefronts of the light and degrade the image quality. Digital aberration correction has been explored using iterative methods in quantitative phase imaging[9,24–27], yet these approaches require parameter tuning and might be sensitive to initial conditions. Although closed-form solution based methods have been introduced for more robust aberration correction, they were demonstrated using only 2D thin samples[28,29]. To date, no technique has successfully integrated spectrum extension, aberration correction, and analytical reconstruction in a single approach for RI tomography.

In this study, we propose Analytic Fourier Ptychotomography (AFP) for 3D complex-valued RI imaging as a way to simultaneously address all the issues discussed above. AFP differs from its predecessor methods by introducing the new concept of using finite sample thickness (FST) prior to expand the sub-spectra (from different illumination angles) overlap. This innovation simplifies the inverse scattering problem into a linear form, allowing aberration correction and darkfield extension to be treated as linear problems that can be solved analytically. Specifically, AFP first reconstructs the complex optical fields using the K-K relation, then analytically corrects aberrations from spectrum overlaps, and finally extends the spectrum beyond the system's physical limits in 3D. AFP enables rapid imaging by eliminating axial scanning, and employs an all-analytic solution that removes the need for manual parameter tuning. It increases space-bandwidth product of the imaging system, achieves large field-of-view (FOV), subcellular resolution, and retains a long working distance. Its simplicity, robustness and efficiency make it a powerful tool for quantitative volumetric RI imaging in weakly scattering samples. We experimentally validated AFP by reconstructing high-resolution 3D images of polystyrene beads under various kind of aberration conditions. Furthermore, we demonstrated its broad applicability across diverse fields, including embryo imaging, plant root and bacteria analysis, and 3D digital pathology.

**Principle**

The AFP pipeline includes image acquisition (Fig. 1a) and analytic RI restoration (Fig. 1b-d). To capture NA-matching and darkfield measurements, AFP uses a standard 4-f microscopy system integrated with multi-angle illumination; it can be implemented using a galvo-mirror scanning system, a light emitting diode (LED) array, or a single LED mounted on a rotational stage (Supplementary Note 2). During image acquisition, the incident planewave from the illumination source interacts with a weakly scattering sample. The total optical field, containing both the unscattered and scattered light, is then captured by the camera and only the intensities are recorded in the image measurement.

To accurately retrieve sample RI distributions from intensity images, two key challenges must be addressed: determining the phase of the detected optical field, and correcting aberrations to restore the sample's optical field accurately. AFP addresses both challenges through a three-step analytic reconstruction process (Fig. 1b-d), which ultimately recovers an aberration-free 3D complex-valued RI distribution.

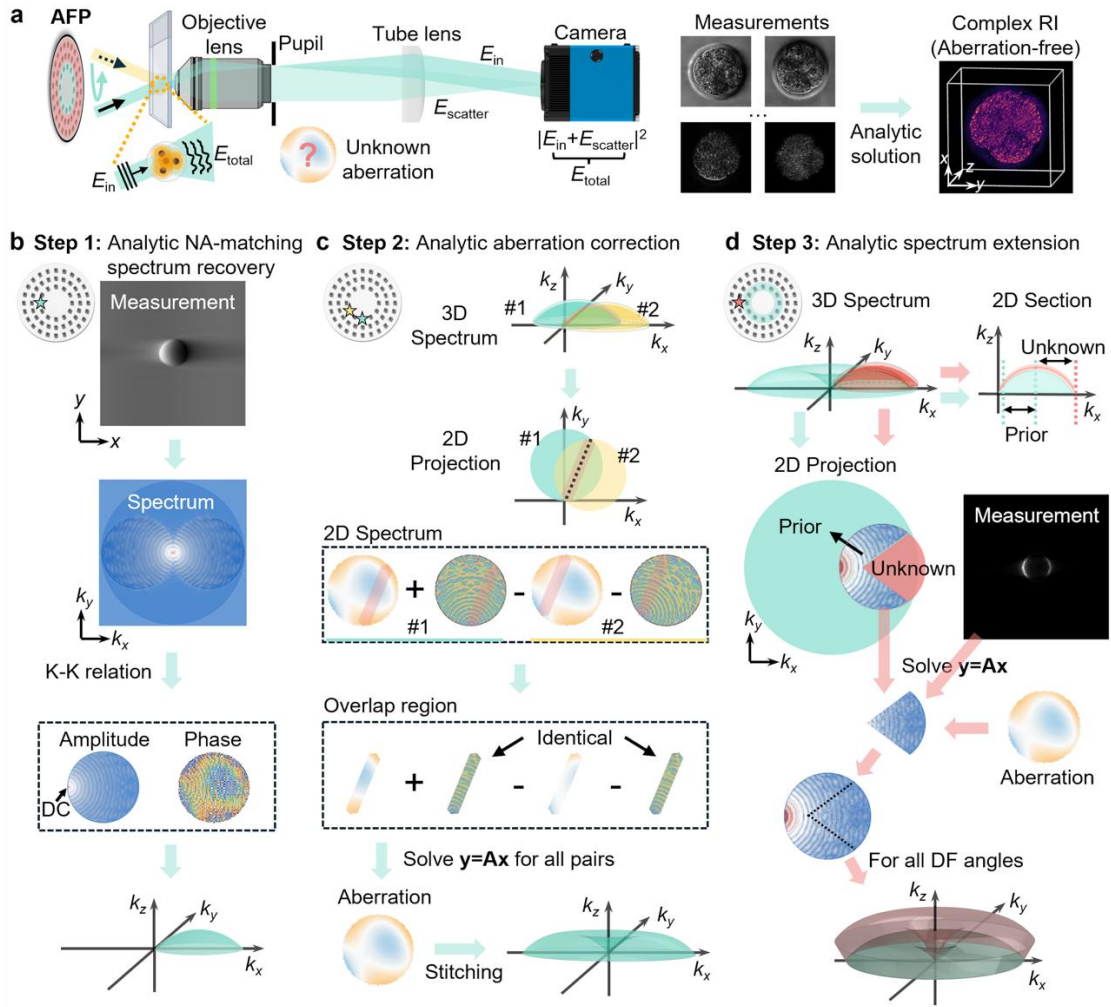

**Fig.1 | Principle and reconstruction pipeline of AFP. a,** Overview of the AFP concept, including the optical setup and analytic reconstruction. Light from different illumination angles interacts with the sample sequentially and gets recorded by a camera. The measurements are then used for linear reconstruction to obtain the complex refractive index (RI) map. **b-d,** Three steps of AFP reconstruction, including analytic NA-matching spectrum (the Fourier transform of the total light field $E_{total}$) recovery (**b**), analytic aberration correction (**c**) and analytic spectrum extension (**d**). In Step 1, Kramers-Kronig (K-K) relation is applied to reconstruct the spectra of the NA-matching measurements. In Step 2, the reconstructed spectra in Step 1 are paired in a group of two and the overlapping part in each pair is extracted for aberration correction. In Step 3, darkfield measurements are sequentially reconstructed: for each sub-step, the corrected spectrum is treated as prior and the unknown part in the spectrum of current darkfield measurement gets linearly solved. Our final reconstruction is obtained after solving for all darkfield measurements. A bead sample is shown as an example in **b-d**. DC indicates the zeroth spatial frequency, DF: darkfield. The insets in the top left corner of **b-d** show the illumination angle design and the active illumination angles used in the corresponding steps. All the spectrum 2D projections specifically refer to the projections onto the $k_x$ - $k_y$ plane.

In Step 1, AFP employs NA-matching measurements to obtain the amplitude of the complex sample fields (Fig. 1b). We then analytically derive the phase of the complex light field using the K-K relation[17,28,30,31]. In a practical imaging system with limited NA, this field corresponds to a small region in the spatial frequency space (and thus will be referred to as a sub-spectrum). Under the weakly scattering assumption[1,32], the first-order scattered field can be directly derived from the reconstructed sub-spectrum. Moreover, Fourier diffraction theorem[1,20] enables the mapping of the scattered field's 2D spectrum onto a 3D spectrum shell constrained by the Ewald sphere (Fig. 1b). This procedure is repeated for each NA-matching illumination angle to reconstruct every sub-spectrum. Since the Fourier transform of a 2D intensity image is the projection of the 3D spectrum shell onto the $k_x$ - $k_y$ plane, AFP works at this plane to reconstruct the complex field associated with each image.

In Step 2, analytic aberration correction is applied to the reconstructed sub-spectra in Step 1 (Fig. 1c). We first consider the case where two 3D sub-spectra from different illumination angles overlap. Theoretically, the two sub-spectra intersect along an arc, and its 2D projection in the $k_x$ - $k_y$ plane is a straight-line segment (black dashed line in Fig. 1c, 2D projection). Given the FST prior, the overlap region of the two sub-spectra can be extended depending on the sample thickness (red-shaded region in Fig. 1c, see details in Supplementary Note 1); the size of the overlap region decreases as the sample thickness increases. Within the overlap region, the sub-spectra from different illumination angles carry the same spatial frequencies of the sample. However, the phases of these sub-spectra differ due to aberrations. By subtracting the phases from these sub-spectra, the phase associated with sample spectrum cancels out, leaving only the aberration differences. It can be shown that these differences contribute to a linear function of the aberration (Supplementary Note 1). By combining these differences in multiple NA-matching measurement pairs, we can assemble a set of linear equations, which can then be solved to recover the system aberration (Supplementary Note 1). The recovered aberration function is then applied to correct all sub-spectra obtained from Step 1. The corrected spectra are then stitched together to produce an aberration-free 3D spectrum from NA-matching measurements.

In Step 3, we analytically extend darkfield spectrum using darkfield measurements under the FST prior (Fig. 1d). A darkfield sub-spectrum does not typically overlap with the 3D spectrum reconstructed in Step 2. However, with the FST prior, part of the darkfield sub-spectrum can be well approximated by the known spectrum from Step 2. It is then used as the prior in the subsequent reconstruction. The rest of the spectrum is treated as an unknown component to be solved (see 2D section and 2D projection in Fig. 1d). The cross-correlation of the prior sub-spectrum and the unknown component corresponds to a linear region in the Fourier transform of a darkfield measurement (Supplementary Note 1). This cross-correlation endows a set of linear equations with respect to the unknown part. Then, we solve the unknown component in an analytic least-squares fashion (see Supplementary Note 1). The darkfield-extended spectrum is then reconstructed by synthesizing all darkfield sub-spectra with the NA-matching spectrum ptychographically. Finally, we obtain the complex-valued volumetric RI distribution by applying a linear transform to the 3D synthetic spectrum (Supplementary Note 1).

## Results

**Validation of AFP with polystyrene beads**

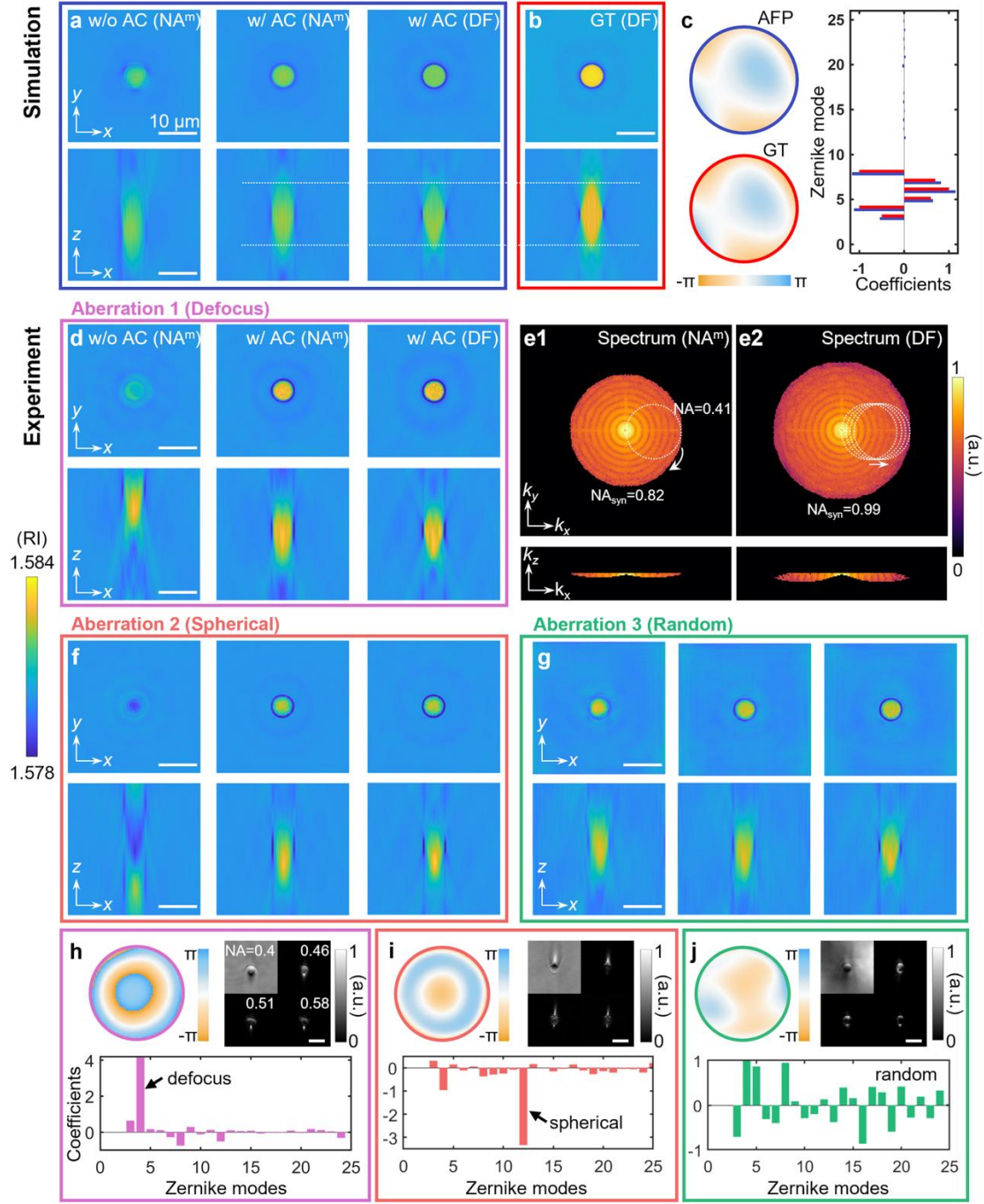

**Fig.2 | AFP with 6-μm polystyrene beads. a-c,** Simulations. **a,** x-y and x-z view comparisons highlight the improvement in image quality from both aberration correction (in shape uniformity) and darkfield extension (in axial resolution). **b,** Ground truth images of the bead with the same darkfield spectrum coverage as in AFP setting shown in panel **a**. **c,** The pupil function and its Zernike decompositions from AFP reconstruction and ground truth profile. **d-j,** Experiments. **d,** x-y and x-z views of AFP reconstruction using a high-quality objective lens (20×/0.41 NA) with a manually induced defocus-dominant aberration. **e,** Visualizations of NA-matching-only (**e1**) and darkfield-extended (**e2**) log-scaled spectrum amplitude from panel **d**, where the ring-shaped spectrum structures are clearly visible. **f,** x-y and x-z views of AFP reconstruction for an invertedly placed sample (spherical-dominant aberration) with the same objective lens in **d**. **g,** x-y and x-z views of AFP reconstruction under a random aberration induced by a piece of scotch tape placed close to pupil plane of the objective lens. **h-j,** Pupil functions and their Zernike decompositions from AFP reconstruction for the three system aberrations in **d, f, and g**, respectively. Examples of the raw intensity measurements are shown in the top-right corner in each panel. GT: ground truth, NA$^m$: NA-matching, DF: darkfield. Scale bar, 10 μm.

We first validated AFP's accuracy in reconstructing the 3D RI profile and correcting system aberration using 6-μm polystyrene beads in both simulations and experiments. In the simulation, a 6-μm diameter bead (n = 1.59) was immersed in a refractive index liquid (n = 1.58), with a known aberration function consisting of six dominant Zernike modes applied to the pupil plane. Without aberration correction, the 3D reconstruction showed severe distortions (Fig. 2a, left), whereas our correction restored the bead's symmetry and uniformity (Fig. 2a, middle). Darkfield extension further improved axial resolution by another twofold compared to NA-matching-only illumination (Fig. 2a, right and Supplementary Note 4). The x-y and x-z sectional profiles demonstrated a high reconstruction quality and were matched with the ground truth obtained with the same 3D spatial frequency support (Fig. 2b). Additionally, the Zernike coefficients of the recovered aberration match the known aberration, verifying the accuracy of aberration correction (Fig. 2c).

In the experiment, we imaged a bead sample using parameters that were matched to those in the simulation (see Methods section). To experimentally evaluate AFP's performance under different aberration conditions, we applied three distinct aberrations: defocus-dominant aberration through sample defocusing, spherical-dominant aberration by inverting slide placement, and random aberrations by attaching scotch tape near the pupil plane of the objective lens. We acquired 48 NA-matching (illumination NA ≈ 0.41) and 144 darkfield images (illumination NA ≈ 0.46, 0.51, 0.58 on three rings) using an LED array-based setup with sequential angular illuminations (see Supplementary Note 2). This configuration extended the synthetic NA from 0.41 (coherent imaging) to 0.82 (NA-matching) and 0.99 (darkfield), resulting in more than a twofold enhancement in lateral resolution.

The reconstructions experimentally show AFP's effective aberration correction (Fig. 2d–j). AFP corrected both single Zernike term dominant aberrations (Fig. 2h,i) and more complex aberrations (Fig. 2j). For example, AFP corrected the defocus-dominant aberration with a maximal phase difference of $2.3\pi$ experimentally. Due to the missing cone problem[33,34] that arises from information loss in z-axis of the 3D Fourier spectrum, the x-z sectional profiles appeared elongated rather than perfectly round. However, with analytic darkfield extension, AFP reduced axial elongation for all aberrations and improved the axial resolution (Fig. 2d,f,g) by a factor of two (see Supplementary Note 4). The 3D spectrum extension was further validated through spectrum visualization, where characteristic ring-shaped structures of the bead spectrum were clearly observed (Fig. 2e).

**Application in label-free tomographic imaging of mouse embryos**

Embryos are optically transparent biological structures that serve as fundamental model organisms for studying early developmental processes, typically ranging from one to several hundred microns in size. Label-free volumetric imaging is crucial for studying and monitoring mouse and human embryo development, particularly in developmental biology and In Vitro Fertilization clinics. Compared to fluorescence microscopy, it minimizes phototoxicity, preserves natural embryo development, and provides high-resolution 3D information non-invasively. In mouse embryos, preimplantation development spans approximately five days, progressing from a single-cell zygote to the two-cell, four-cell, eight-cell, and ultimately the blastocyst stage[35,36]. As the development advances, cellular divisions create increasingly sophisticated 3D structures. In Fig. 3, we present AFP results from fixed mouse embryos at the four-cell and eight-cell stages, acquired using a laser-based setup (see Supplementary Note 2). Additional results and visualizations of fixed and live mouse embryos are provided in the Supplementary Videos 1,2 and Supplementary Note 9.

We showed 3D maximum intensity projection (MIP) renderings of whole embryos, with individual nucleus segmentations overlaid (Fig. 3a,c). The quantitative RI reconstruction provides high-contrast visualization of the nuclear membrane boundary and nucleoli inside each cell, facilitating calculations of nucleus to cell boundary distance and the center of mass that are keys to evaluating embryo early development (Fig. 3b)[37,38]. The 3D structure across multiple z-planes reveals the spatial distribution of nucleoli, which could aid nucleus tracking during the division processes. For the eight-cell stage embryo, the eight cells are distinctly distributed across different z-planes, with all eight nuclei clearly visible in three representative z-sections (Fig. 3d). With darkfield extension, finer intracellular structures become more discernible as shown in the x-z sectional profile in Fig. 3e. Finally, we compared the AFP image of the eight-cell stage mouse embryo with the image from absorption-based widefield microscopy (Fig. 3f and Supplementary Video 3). AFP provides a clearer view of cell structures, particularly enabling a more comprehensive analysis of the nucleoli. In contrast, absorption-based microscopy suffers from low contrast in transparent samples, leading to information loss regarding the cell's internal features.

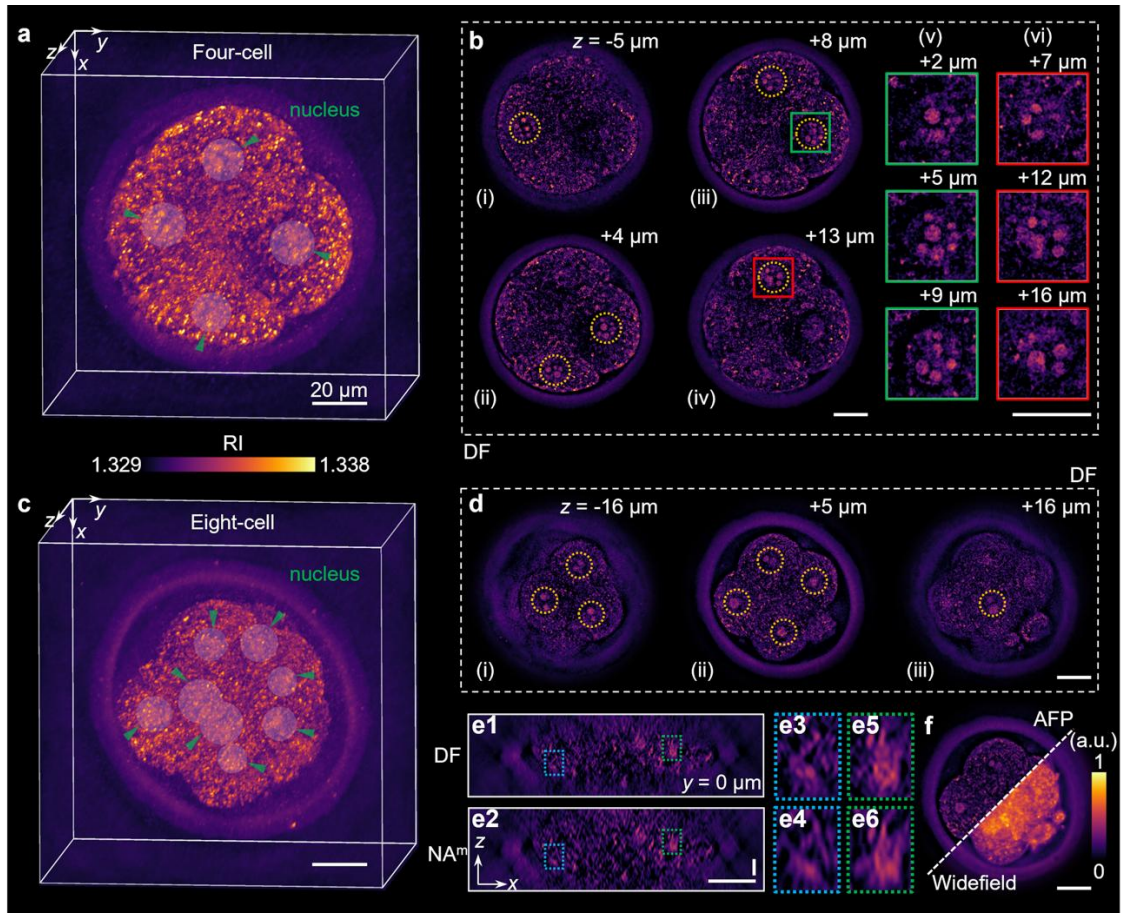

**Fig.3 | Label-free RI tomography of fixed mouse embryos with AFP. a,b,** A four-cell stage embryo imaged with AFP. **a,** 3D MIP rendering of the AFP reconstruction. The individual nucleus segmentations are pointed by green arrows. **b,** Visualization of four z-slices showing nucleoli within each embryonic cell (i-iv), along with two enlarged views of nucleoli distributions across three distinct z-slices (v-vi). **c-f,** An eight-cell stage embryo imaged with AFP. **c,** 3D MIP rendering of the AFP reconstruction. Green arrows indicate nucleus segmentations overlaid. **d,** Eight nuclei within each cell are marked in three representative z-slices. **e,** Comparison of x-z views between darkfield (**e1**) and NA-matching (**e2**) AFP reconstructions, with magnified views (blue and green boxes) shown in **e3-e6**. **f,** Imaging comparison between AFP and an absorption-based widefield microscope at z = +5 μm. NA$^m$: NA-matching, DF: darkfield. Scale bar, 20 μm.

## Application in large FOV plant root and bacteria imaging

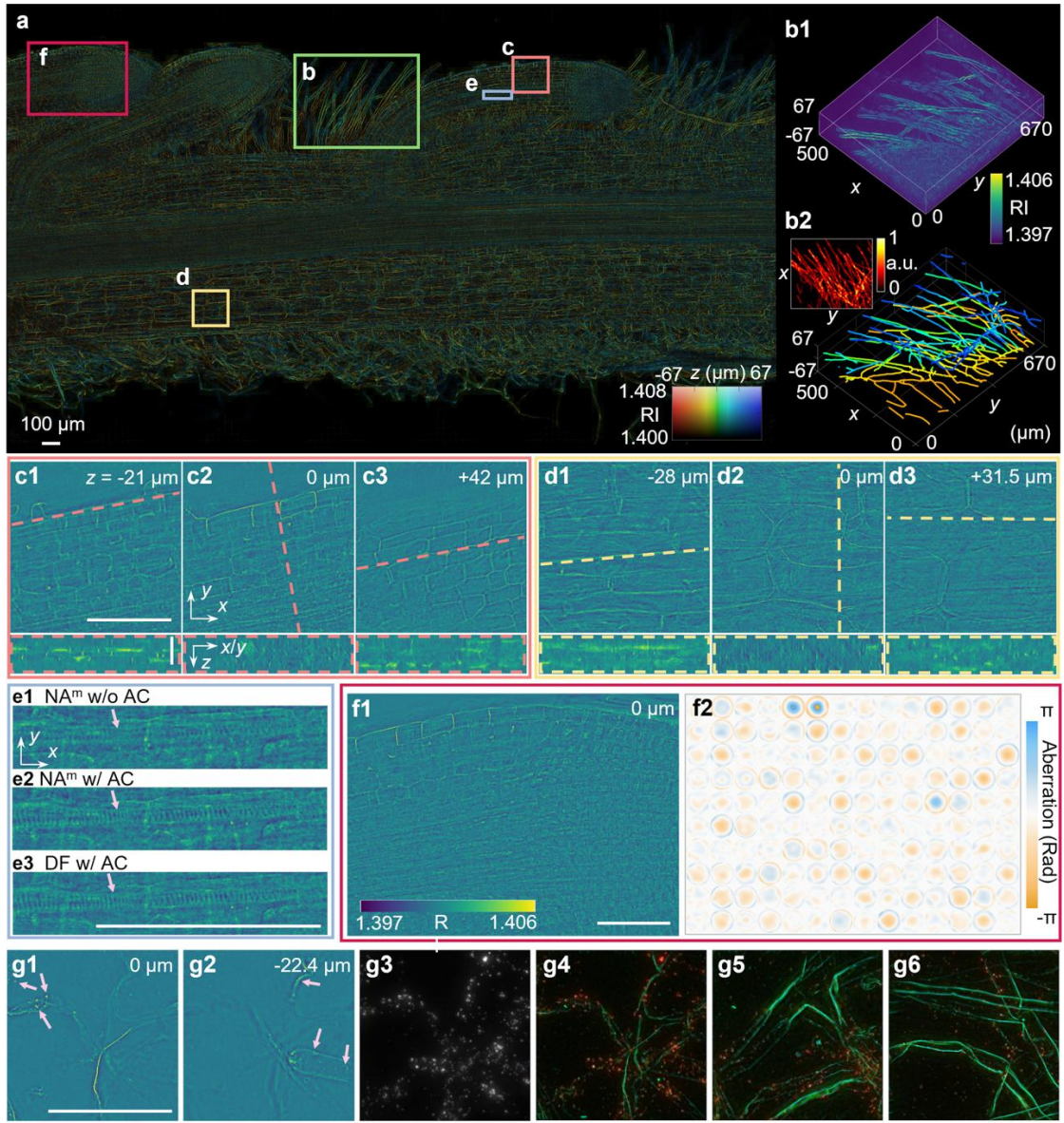

**Fig.4 | Wheat root and bacteria imaging with AFP. a,** MIP of a 4.18×2.40×0.140 mm$^3$ volume RI image using AFP with depth encoded in colors and RI encoded in image brightness. **b1,** A 3D MIP rendering of root hair region in **a**. **b2,** 3D root skeleton extracted from **b1** and associated root hair density map at top-left corner. **c,d,** Different z-slices of cellular structures from cortex regions in **a**. **e,** aberration correction and darkfield extension provides a clear structure of the stele region from a baby root tip in **a**. **f,** Field-dependent aberration correction for different patches of root structures. The right panel shows the retrieved aberrations. **g,** AFP for imaging the coexistence of root hair and bacteria (see arrows) in a 140$^3$ μm$^3$ volume (**g1,g2**). **g3,** MIP of fluorescence images for bacteria in the same FOV as **g1** and **g2**. **g4,** Overlayed RI (in green) and fluorescence (in red) images in the same FOV as **g1**. **g5,g6,** More FOVs for overlayed root hair and bacteria imaging. NA$^m$: NA-matching, AC: aberration correction, w/o: without, w/: with, DF: darkfield. Scale bar, 100 μm.

The rhizosphere is a dynamic interface between plant roots and the soil microenvironment, where critical biological interactions occur. Visualizing root structure is essential for understanding root development, function, and interactions with the soil environment, with volumetric information providing crucial insights into root morphology, spatial organization, and root-soil interactions in three dimensions[39]. Here, we present the RI reconstruction of a 7-

day-old wheat root section (*Triticum aestivum*) over a 4.18×2.40×0.140 mm$^3$ volume (Fig. 4a), obtained using an LED-based microscope system with 48 NA-matching (illumination NA = 0.40) and 288 darkfield illumination angles (synthetic NA = 0.98). The RI distribution enables high-resolution visualization of various root structures. Root hairs, which facilitate water and nutrient uptake, appear as elongated cells (Fig. 4b, Supplementary Video 4). From 3D RI distribution, we extracted the root hair skeleton, allowing for spatial density and heterogeneity analysis (Fig. 4b, Supplementary Video 5). Additionally, epidermis, cortex and endodermis cells at the primary root or lateral root, layered around vascular bundle, are distinctly visualized (Fig. 4c,d, Supplementary Video 4). Despite the rich volumetric structural information, our aberration correction approach effectively compensates for field-dependent aberrations, enabling the recovery of fine cellular and subcellular structures (Fig. 4e,f).

In addition to root structures, *Pseudomonas synxantha* 2-79, a rhizosphere bacteria which contribute to plant growth and disease suppression, was also imaged alongside root hairs (Fig. 4g, Supplementary Note 10). Conventionally, visualizing the interaction between bacteria and root structures requires labeling of both and imaging with fluorescence microscopy. AFP, owing to its label-free nature, do not suffer from low signal-level and photobleaching during repeated z-scanning, can accomplish low noise and fast acquisition. These advantages enable comprehensive reconstructions of root and bacteria structures in 3D, making AFP a highly suitable imaging method for rhizosphere research and environmental studies.

**Complex-valued RI imaging for 3D digital pathology**

Digital pathology analysis has become a cornerstone in cancer diagnosis and patient treatments. Recent advancements in 3D digital pathology and deep learning suggest that the spatial volumetric information in histology can significantly enhance tumor grading and diagnosis[40]. Here, we examine AFP operating at a deep ultra-violet (DUV) wavelength for 3D label-free histologic imaging, eliminating variations from preanalytical sample preparation factors. The use of DUV in this application is motivated by the strong absorption exhibited by nucleic acids at this wavelength.

Conventional histology relies on hematoxylin and eosin (H&E) stains to label the nucleic acids and extracellular matrix. However, H&E stain often requires complex preparation procedures for thick tissue samples[41]. Additionally, variability in the stain quality and color consistency poses challenges for deep learning-based analysis[42]. To address these limitations, the DUV AFP leverages complex-valued RI distributions at DUV wavelength to mimic H&E stain contrast.

We demonstrated RI imaging of a 20-μm-thick human gastric adenocarcinoma section (Fig. 5) using DUV illumination at 265 nm with 48 NA-matching measurements (See Supplementary Note 2). The real part of the RI distributions modifies the optical path length through the sample, reflecting variations in tissue composition and revealing intricate histologic features (Fig. 5a). The imaginary part of the RI distribution correlates with tissue absorptance (See Supplementary Note 11), as nucleic acids strongly absorb light at 260–280 nm wavelengths due to π-electron transitions in aromatic bases[43]. Consequently, the nuclear structures are clearly resolved in the absorption image (Fig. 5b), effectively replicating the functionality of H-stain. Furthermore, other z-slices of the reconstructed complex-valued RI volume are presented in Fig. 5c-f (see more results in Supplementary Video 6 and Supplementary Note 11). To directly compare DUV AFP with traditional H&E staining results, we imaged a H&E-stained, 5-μm thin section from an adjacent cut under a 20×/0.40 NA widefield microscope (Fig. 5g). Additionally, we showed that dividing the entire FOV into small patches allowed for correction of field-dependent aberrations and enhanced image quality (Fig. 5h-k, Supplementary Video 6). These results demonstrate AFP's ability to correct image blur caused by aberrations in real-world applications.

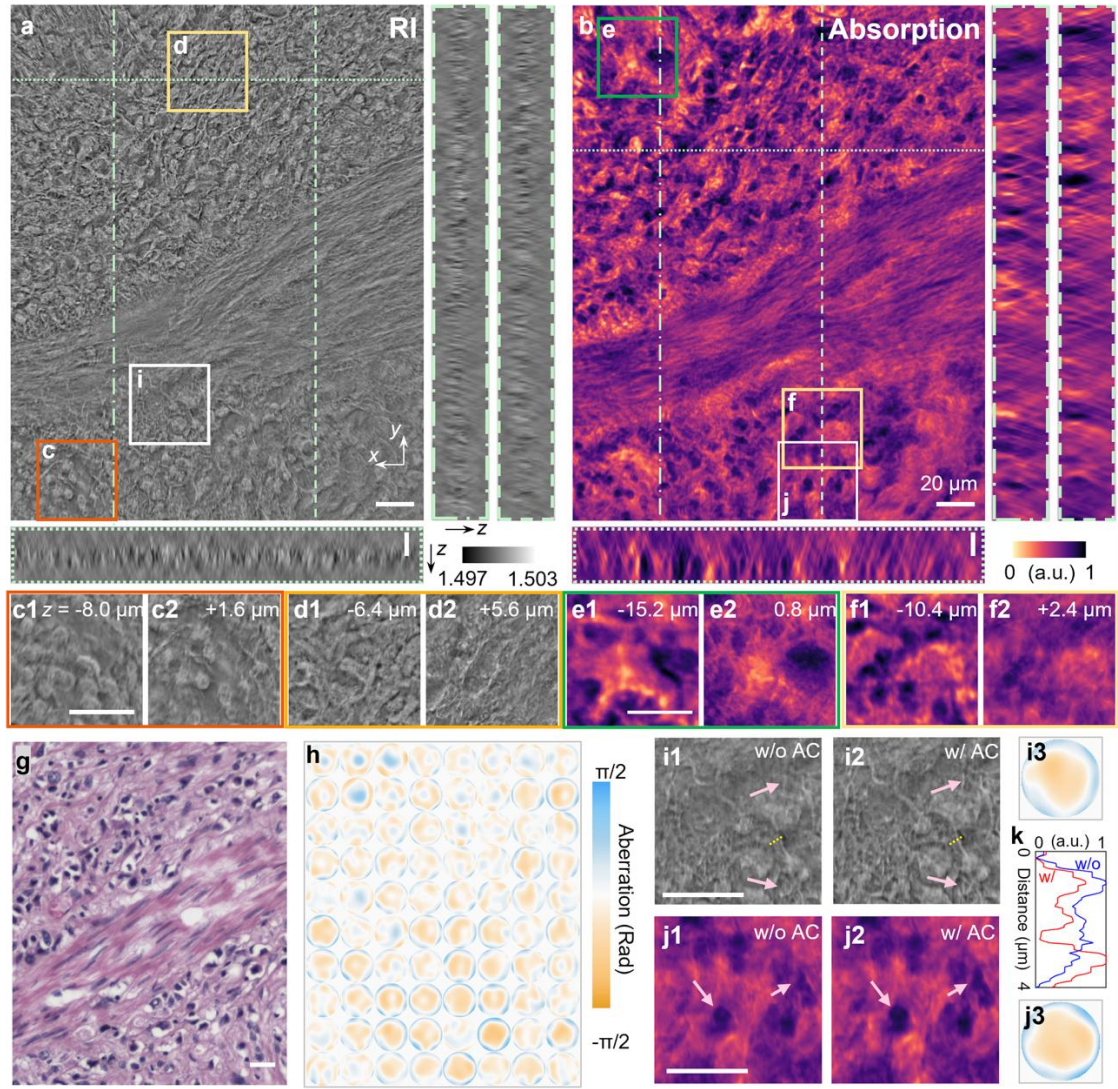

**Fig.5 | 20-μm thick human gastric adenocarcinoma cancer pathological slide imaging with DUV AFP. a,** A z-slice of the RI real part volume from AFP with x-z and y-z sections. **b,** The absorption image correlated with the imaginary part of RI from AFP with associated x-z and y-z views. **c-f,** z-slices of zoomed-in views in (**a,b**). **g,** The widefield microscope image of an H&E-Stained adjacent thin cut. **h,** Retrieved aberrations from (**a,b**), indicating effective field-dependent aberration correction. **i,j,** Examples of image contrast improvements with the aberration correction. **k,** Line profiles comparison between yellow dashed lines in **i1** and **i2**. AC: aberration correction, w/o: without, w/: with. Scale bar, 20 μm.

## Discussion

In summary, we have demonstrated AFP, a technique that transforms intensity measurements from multiple angular illuminations into aberration-free, complex-valued 3D RI distributions. Unlike many other computational microscopy methods[5,6,19,20], AFP's analytic formulation eliminates the need for parameter tuning (see Supplementary Note 7), making it highly robust and generalizable across different sample types and system configurations. We demonstrated the versatility of AFP through experiments on a variety of samples and setups, including polystyrene beads, mouse embryos, wheat roots and human gastric tissues using multiple microscope systems with various illumination sources (lasers and LEDs spanning visible to DUV wavelengths, see additional samples in Supplementary Notes 8-11). Further, it extends the resolution beyond the NA of the objective lens through linear 3D spectrum extension, while preserving long working distances when using low-magnification objective lenses. These features make AFP widely applicable for imaging large samples at high resolution. For example, a promising direction to explore is to combine AFP with expansion microscopy, where cellular contents, distantly embedded from the coverslip, are impossible to image due to physically expanded tissues[44].

The ability of AFP to map 3D RI distribution makes it well-suited for quantitative analysis of weak scattering microscopic samples. For example, embryo imaging using AFP highlights its potential in providing detailed insights into cell morphology, size, location, dry mass, and structural organization. Additionally, AFP's visualization of root structures mixed with rhizosphere bacteria aligns excellently with fluorescence imaging, yet without the inherent limitations of fluorescence methods such as photobleaching. Further, DUV AFP imaging of pathology slides provides volumetric images that are independent of staining processes, which can vary across different sample preparation facilities and stain concentrations for conventional histology. We anticipate that this quantitative RI reconstruction, combined with deep learning techniques, can enable a range of downstream applications, such as embryo health assessment at early developmental stages, automatic differentiation of bacteria from root structures based on morphological features, label-free virtual staining of pathological slides, and cancer prognosis and diagnosis.

AFP also holds good potential for live cell and organoid imaging application, because of its rapid data acquisition and robust image reconstruction advantages. Currently, using a laser-based system, AFP completes an acquisition of 16-bit 600×600×366-voxels dataset in 42 seconds (for embryo sample), with NA-matching-only captures taking 3.8 seconds. Additionally, there remains room for further acceleration. For example, the use of brighter light sources, faster cameras, hardware triggering for synchronization, and a higher data transmission bandwidth will reduce the acquisition time. Moreover, tailoring illumination angles for specific applications could minimize data redundancy and improve imaging speed. For image reconstruction, our current algorithm takes approximately 6 minutes to reconstruct a volume of 480×480×400 voxels using an AMD EPYC 7763 CPU and an Nvidia A100-80 GB GPU (for wheat root sample). Parallel computing with two CPUs and four GPUs of the same specifications can further reduce this processing time to just one minute. Further acceleration could be achieved through PyTorch-based implementations[45], which may improve computational efficiency via programming[29].

A current limitation of AFP is that it has so far been validated only in the weakly scattering regime, where the single scattered light dominates over the multiple-scattered light. In our experiments, we measured that the scattered light intensity from the sample is less than 25% of the incident light intensity. Specifically, our experimental samples each have an average thickness of less than 100 μm and a sample-background RI difference of less than 0.01. Currently, AFP does not account for multiple scattering effects in optically thick samples or those with significant RI differences relative to the background. Addressing these conditions still requires the use of optimization-based methods, which nonetheless suffer from the high computational complexity and non-convexity[10,12,14]. However, AFP could potentially serve as an effective initialization for these optimization approaches to tackle aberrations, improve convergence speed, and decrease the likelihood of settling at suboptimal solutions. Further, integrating AFP initialization with appropriate regularization techniques, such as sparsity constraints and non-negative constraints, can further enhance image quality and mitigate artifacts associated with the missing-cone problem.

We envision that AFP's simple hardware design and robust algorithm will catalyze high-throughput applications for volumetric RI imaging across various fields, including cell biology, in vitro embryo development study, 3D digital pathology, tissue engineering, and plant science. Moreover, the label-free and quantitative imaging capabilities will facilitate biological and clinical data analysis, particularly in the context of machine learning and deep learning. This advancement is also expected to inspire innovative designs of imaging systems, for example, for addressing multiple-scattering challenges and enhance overall image quality.

# References

1. Wolf, E. Three-dimensional structure determination of semi-transparent objects from holographic data. *Opt. Commun.* **1**, 153–156 (1969).
2. Kim, G. *et al.* Holotomography. *Nat. Rev. Methods Primer* **4**, 1–22 (2024).
3. Sung, Y. *et al.* Optical diffraction tomography for high resolution live cell imaging. *Opt. Express* **17**, 266–277 (2009).
4. Kim, K. *et al.* High-resolution three-dimensional imaging of red blood cells parasitized by Plasmodium falciparum and in situ hemozoin crystals using optical diffraction tomography. *J. Biomed. Opt.* **19**, 011005 (2013).
5. Chen, M., Tian, L. & Waller, L. 3D differential phase contrast microscopy. *Biomed. Opt. Express* **7**, 3940–3950 (2016).
6. Li, J. *et al.* High-speed in vitro intensity diffraction tomography. *Adv. Photonics* **1**, 066004 (2019).
7. Zhao, J. *et al.* Bond-selective intensity diffraction tomography. *Nat. Commun.* **13**, 7767 (2022).
8. Hugonnet, H., Lee, M. & Park, Y. Optimizing illumination in three-dimensional deconvolution microscopy for accurate refractive index tomography. *Opt. Express* **29**, 6293–6301 (2021).
9. Chung, Y., Hugonnet, H., Hong, S.-M. & Park, Y. Fourier space aberration correction for high resolution refractive index imaging using incoherent light. *Opt. Express* **32**, 18790–18799 (2024).
10. Chowdhury, S. *et al.* High-resolution 3D refractive index microscopy of multiple-scattering samples from intensity images. *Optica* **6**, 1211–1219 (2019).
11. Tian, L. & Waller, L. 3D intensity and phase imaging from light field measurements in an LED array microscope. *Optica* **2**, 104–111 (2015).
12. Chen, M., Ren, D., Liu, H.-Y., Chowdhury, S. & Waller, L. Multi-layer Born multiple-scattering model for 3D phase microscopy. *Optica* **7**, 394–403 (2020).
13. Zhu, J., Wang, H. & Tian, L. High-fidelity intensity diffraction tomography with a non-paraxial multiple-scattering model. *Opt. Express* **30**, 32808–32821 (2022).
14. Lee, M., Hugonnet, H. & Park, Y. Inverse problem solver for multiple light scattering using modified Born series. *Optica* **9**, 177–182 (2022).
15. Liu, H.-Y. *et al.* SEAGLE: Sparsity-Driven Image Reconstruction Under Multiple Scattering. *IEEE Trans. Comput. Imaging* **4**, 73–86 (2018).
16. Tong, Z. *et al.* Three-dimensional refractive index microscopy based on the multi-layer propagation model with obliquity factor correction. *Opt. Lasers Eng.* **174**, 107966 (2024).
17. Baek, Y. & Park, Y. Intensity-based holographic imaging via space-domain Kramers–Kronig relations. *Nat. Photonics* **15**, 354–360 (2021).
18. Li, J. *et al.* Transport of intensity diffraction tomography with non-interferometric synthetic aperture for three-dimensional label-free microscopy. *Light Sci. Appl.* **11**, 154 (2022).
19. Horstmeyer, R., Chung, J., Ou, X., Zheng, G. & Yang, C. Diffraction tomography with Fourier ptychography. *Optica* **3**, 827–835 (2016).
20. Zuo, C., Sun, J., Li, J., Asundi, A. & Chen, Q. Wide-field high-resolution 3D microscopy with Fourier ptychographic diffraction tomography. *Opt. Lasers Eng.* **128**, 106003 (2020).
21. Wu, X. *et al.* Lens-free on-chip 3D microscopy based on wavelength-scanning Fourier ptychographic diffraction tomography. *Light Sci. Appl.* **13**, 237 (2024).
22. Zheng, G., Horstmeyer, R. & Yang, C. Wide-field, high-resolution Fourier ptychographic microscopy. *Nat. Photonics* **7**, 739–745 (2013).
23. Zheng, G., Shen, C., Jiang, S., Song, P. & Yang, C. Concept, implementations and applications of Fourier ptychography. *Nat. Rev. Phys.* **3**, 207–223 (2021).

24. Ou, X., Zheng, G. & Yang, C. Embedded pupil function recovery for Fourier ptychographic microscopy. *Opt. Express* **22**, 4960–4972 (2014).
25. Tian, L., Li, X., Ramchandran, K. & Waller, L. Multiplexed coded illumination for Fourier Ptychography with an LED array microscope. *Biomed. Opt. Express* **5**, 2376–2389 (2014).
26. Chen, M., Phillips, Z. F. & Waller, L. Quantitative differential phase contrast (DPC) microscopy with computational aberration correction. *Opt. Express* **26**, 32888–32899 (2018).
27. Huang, Z., Yang, F., Liu, B., Liu, Y. & Cao, L. Aberration-free synthetic aperture phase microscopy based on alternating direction method. *Opt. Lasers Eng.* **160**, 107301 (2023).
28. Cao, R., Shen, C. & Yang, C. High-resolution, large field-of-view label-free imaging via aberration-corrected, closed-form complex field reconstruction. *Nat. Commun.* **15**, 4713 (2024).
29. Zhao, S., Zhou, H., Lin, S. (Steven), Cao, R. & Yang, C. Efficient, gigapixel-scale, aberration-free whole slide scanner using angular ptychographic imaging with closed-form solution. *Biomed. Opt. Express* **15**, 5739–5755 (2024).
30. Shen, C., Liang, M., Pan, A. & Yang, C. Non-iterative complex wave-field reconstruction based on Kramers–Kronig relations. *Photonics Res.* **9**, 1003–1012 (2021).
31. Baek, Y., Lee, K., Shin, S. & Park, Y. Kramers–Kronig holographic imaging for high-space-bandwidth product. *Optica* **6**, 45–51 (2019).
32. Devaney, A. J. Inverse-scattering theory within the Rytov approximation. *Opt. Lett.* **6**, 374–376 (1981).
33. Lee, M., Shin, S. & Park, Y. Reconstructions of refractive index tomograms via a discrete algebraic reconstruction technique. *Opt. Express* **25**, 27415–27430 (2017).
34. Lim, J., Ayoub, A. B. & Psaltis, D. Three-dimensional tomography of red blood cells using deep learning. *Adv. Photonics* **2**, 026001 (2020).
35. Shen, C. *et al.* Stain-free detection of embryo polarization using deep learning. *Sci. Rep.* **12**, 2404 (2022).
36. Goolam, M. *et al.* Heterogeneity in Oct4 and Sox2 Targets Biases Cell Fate in 4-Cell Mouse Embryos. *Cell* **165**, 61–74 (2016).
37. Junyent, S. *et al.* The first two blastomeres contribute unequally to the human embryo. *Cell* **187**, 2838-2854.e17 (2024).
38. Chen, Q., Shi, J., Tao, Y. & Zernicka-Goetz, M. Tracing the origin of heterogeneity and symmetry breaking in the early mammalian embryo. *Nat. Commun.* **9**, 1819 (2018).
39. Zhang, O. *et al.* Investigating 3D microbial community dynamics of the rhizosphere using quantitative phase and fluorescence microscopy. *Proc. Natl. Acad. Sci.* **121**, e2403122121 (2024).
40. Song, A. H. *et al.* Analysis of 3D pathology samples using weakly supervised AI. *Cell* **187**, 2502-2520.e17 (2024).
41. Li, Y. *et al.* Hematoxylin and eosin staining of intact tissues via delipidation and ultrasound. *Sci. Rep.* **8**, 12259 (2018).
42. Lin, S. *et al.* Impact of stain variation and color normalization for prognostic predictions in pathology. *Sci. Rep.* **15**, 2369 (2025).
43. Tataurov, A. V., You, Y. & Owczarzy, R. Predicting ultraviolet spectrum of single stranded and double stranded deoxyribonucleic acids. *Biophys. Chem.* **133**, 66–70 (2008).
44. Kubalová, I. *et al.* Prospects and limitations of expansion microscopy in chromatin ultrastructure determination. *Chromosome Res.* **28**, 355–368 (2020).
45. Paszke, A. *et al.* PyTorch: An Imperative Style, High-Performance Deep Learning Library. Preprint at https://doi.org/10.48550/arXiv.1912.01703 (2019).

## Methods

### Imaging systems

Three imaging systems were implemented in this study and are listed below. More detailed descriptions of the imaging systems design and calibration can be found in Supplementary Notes 2 and 3.

The laser-based imaging system utilized a 532 nm continuous-wave laser (CrystaLaser Inc.), which was collimated and then raster scanned by a galvo mirror pair (Thorlabs GVS 212) to vary the incident angles. The optical field was captured by a 4-f imaging system comprising an objective lens (Olympus Plan N 20×/0.40 NA), a tube lens (Thorlabs TTL180-A) and a camera (Excelitas, PCO Edge 5.5). This system was used for imaging mouse embryos and the algae sample (Supplementary Video 7, Supplementary Note 8).

The LED array imaging system featured a customized, programmable LED array (central wavelength: 520 nm) as the light source, a motorized translation stage (Thorlabs MLS203-1) for large FOV scanning, and a detection system built upon a commercial olympus IX51 inverted microscope with an objective lens (Olympus Plan N 20×/0.40 NA) and a camera (Allied Vision Prosilica GT6400). This system was employed for bead imaging as well as root and bacteria imaging experiments.

The DUV system consisted of a 265 nm LED (Boston Electronics) mounted on a motorized rotational stage (Thorlabs ELL14) for illumination, paired with a 20×/0.37 NA objective lens (Mitutoyo Plan UV Infinity Corrected Objective) and a tube lens (Thorlabs TTL200-UVB) and a camera (Allied Vision, Alvium 1800 U-812 UV) for transmitted light detection. This system was used for imaging the human gastric adenocarcinoma cancer cells.

### Simulations

The simulations included reconstructions of a polystyrene bead and a lymph node sample. Specifically, a 6-μm diameter spherical bead sample (n = 1.59) was immersed in a refractive index liquid (n = 1.58), and the lymph node sample (n = 1.33~1.34) was immersed in water (n = 1.33). We simulated the raw measurements under NA-matching and darkfield illuminations using the multi-slice beam propagation method. Various levels of aberration were digitally introduced at the Fourier plane of the imaging system, and the corresponding correction results are shown in Supplementary Note 5. AFP reconstruction enabled the generation of a synthetic 3D spectrum mask, corresponding to the ptychographic sampling of Ewald sphere combining different illumination angles. The ground truth images under NA-matching and darkfield extension (Fig. 2b) were obtained by applying this low-pass 3D spectrum mask to the 3D scattering potential spectrum of a spherical bead and converting it back to the refractive index. The origin of these 3D spectrum masks are explained in Supplementary Notes 4 and 6.

### Sample preparation

**Polystyrene beads.** A standard glass slide (1 mm thickness) and cover glass (0.13 - 0.17 mm thickness) were rinsed with deionized water, wiped with acetone solution, dried with air blow, and then cleaned using a plasma cleaner to remove any small particles. The 6-μm diameter plain polystyrene bead (Abvigen Inc., n ≈ 1.59) solution was diluted 1:200 with 100% Ethanol. 10 μL of the diluted bead solution was applied to the glass slide and allowed to air dry for 10 minutes. Next, another 10 μL of refractive index liquid (Cargille, n ≈ 1.58) was applied to the center of the sample slide and allowed to stand for 10 minutes. Finally, the cover glass was placed on top of the sample and sealed with nail polish.

**Mouse embryos.** All animal works were approved by the IACUC. Either 6-week ages of C57BL/6J or B6D2F1/J female mice (Jackson Laboratory) were used for this study. Female mice were super-ovulated by post-pregnant mare's serum gonadotrophin (PMSG) and human chorionic gonadotrophin (hCG) injection. Zygotes were collected in M2 medium 22 hours after hCG injection and subsequently cultured in Advanced KSOM medium at 37 °C with 5% CO2 in the air. Either live or fixed embryos were used for this study. Embryos were fixed with 4 % paraformaldehyde (PFA) in PBS for 10 min at room temperature (25 °C). Fixed embryos were washed three times with 1% BSA in PBS and kept at 4 °C until further analysis. The fixed embryos were placed on a glass slide with 7 μL of 1% BSA in PBS (n ≈ 1.33), while live embryos were transferred in 7 μL of KSOM solution (n ≈ 1.33) for imaging. The embryos were placed in sperate wells of adhesive silicone isolators (9 mm in diameter and 0.8 mm in depth), covered with a coverslip and sealed with nail polish. Two-cell, Four-cell, Eight-cell and blastocyst embryos were used for this study.

**Wheat root and bacteria preparation and growth conditions.** Soft white wheat seeds (*Triticum aestivum*; Handy Pantry, Lot 190698) were surface sterilized by incubating with a 50% (v/v) bleach solution containing 0.1% (v/v) Triton X for 3 minutes and rinsed five times with autoclaved water. Sterilized seeds were plated on 0.6% phytagel (Sigma Aldrich, Cat. No. P8169) containing 0.5× Murashige and Skoog (MS) medium (Sigma Aldrich, Cat. No. M5519) at pH 5.8 and incubated in a growth chamber under a 16-hour light/8-hour dark cycle at 25°C for 7 days. On day 7, seedlings were collected and cleared by incubating in 1M KOH overnight, followed by rinsing with 0.1M HCl and deionized water before imaging. *Pseudomonas synxantha* $P_{PA10403}$-mNeonGreen (a constitutively fluorescent strain previously constructed) was cultured overnight in LB broth (Sigma-Aldrich, Cat. No. L3522) at 30°C. Bacterial cells were washed three times in PBS buffer and diluted to an optical density of 0.7. Cleared wheat roots were immersed in this bacterial suspension for 30 minutes prior being mounted for imaging. A 1:1 mixture of 100% Glycerol and deionized water was used as mounting media (n ≈ 1.40).

**3D Pathology slides.** Human gastric adenocarcinoma sections were prepared by Acepix Biosciences Inc. The 20-μm-thick sections for DUV imaging were cut from formalin-fixed paraffin-embedded (FFPE) tissue blocks. The samples were prepared on quartz coverslips and slides (Technical Glass Products, Inc.) with Micromount mounting media (Leica Biosystems Inc., n ≈ 1.498), but without any H&E staining. The rest processes were the same as standard FFPE histology slide preparations. The 5-μm adjacent thin cuts were prepared by Acepix Biosciences Inc. with standard H&E-stained histology protocol.

## Experiment details

In the bead experiment, the illumination angles of the LED array setup were pre-calibrated using a 2D Siemens star target (see Supplementary Note 3 for calibration details). To reduce background noise and achieve high-quality darkfield reconstruction, raw darkfield measurements were corrected before reconstruction by subtracting images taken with the pure refractive index liquid. This preprocessing step is particularly crucial for imaging thin samples with a small RI difference compared to the background medium, as such samples produce weak signals under darkfield illumination.

## Computing platform and units

All code running was done by MATLAB (Mathworks, Inc. versions 2021b – 2024b) on either a Windows 10/11 or Linux Ubuntu 22.04 platform. For reference, the code has been tested using Intel i7-12900KF, AMD Ryzen 9 5950X, AMD EPYC 7763 CPUs, and/or Nvidia RTX 3090, 4080Ti, A100-80GB GPUs, while GPU computing is optional.

## Data availability

The data that support this study are publicly available at https://osf.io/f7tqa/

## Code availability

The code for AFP reconstruction is available at https://github.com/MrDongZhenyu/AFP

## Acknowledgements

This research is supported by Rothenberg Innovation Initiative (RI$^2$) (Award number A4188-Yang-3-A1) and the Heritage Medical Research Institute (HMRI) (Award number HMRI-15-09-01). The authors thank Siyuan Yin for helping LED array control, Dr. Shoma Nakagawa for preparing the mouse embryo samples, and Dr. Magdalena Zernicka-Goetz for insightful discussion of the embryo imaging results.

## Author contributions

R.C. conceived the idea. R.C., Z.D. and H.Z. initiate the project. R.C., Z.D. and H.Z. conducted simulations and wrote the reconstruction algorithms. R.C. conducted the theoretical analysis. Z.D. and H.Z. built experimental setups and performed the calibration of the system. Z.D. and H.Z. conducted the experiments with the help of O.Z. and S.Z. Z.D. and H.Z. wrote the reconstruction codes for the experiments. P.L. designed and fabricated the LED array. R.A. provided the wheat root and bacteria sample and proposed the biological insights. Z.D., H.Z., O.Z. and R.C. analyzed the data and performed visualizations. C.Y. supervised this project. All authors contributed to the preparation of the manuscript.

## Competing interests

The authors declare the following competing interests: On April 1, 2025, California Institute of Technology filed a provisional patent application for AFP, which covered the concept and implementation of the AFP method described here.

## Supplementary information

A Supplementary information file and Supplementary Videos 1-7 are attached to this manuscript.